\documentclass[%
 reprint,
 amsmath,amssymb,
 aps,
]{revtex4-1}
\usepackage{graphicx}
\usepackage{epstopdf}
\usepackage{dcolumn}
\usepackage{bm}
\usepackage{balance}

\begin{document}


\title{Resonant high-energy bremsstrahlung of ultrarelativistic electrons in the field of a nucleus and an electromagnetic wave}

\author{Alexander Dubov}
 \email{alexanderpolytech@gmail.com}
\affiliation{
 Department of Theoretical Physics Peter the Great St. Petersburg Polytechnic University\\
St-Petersburg, Russia}

\author{Victor V. Dubov}
 \email{dubov@spbstu.ru}
 \affiliation{
 Department of Theoretical Physics Peter the Great St. Petersburg Polytechnic University\\
 St-Petersburg, Russia}

\author{Sergei P. Roshchupkin}
 \email{serg9rsp@gmail.com}
\affiliation{
 Department of Theoretical Physics Peter the Great St. Petersburg Polytechnic University\\
 St-Petersburg, Russia}

\date{\today}

\begin{abstract}
The actual theoretical research investigates the resonant spontaneous bremsstrahlung (RSB) of ultrarelativistic electrons under the condition of scattering on a nucleus in the field of a weak electromagnetic wave. The progression of the functional mechanism indicates the transformation of the intermediate virtual electron into the real particle state. As a result, the initial second order process with accordance to the fine structure constant in the light field productively splits into two consequent first order formations: the laser-stimulated Compton effect and the laser-assisted scattering of an electron on a nucleus. Precise examination specifies the resonant kinematics of the RSB system that designate the phenomenon of the initial and final electrons with addendum of spontaneous high-energy photon propagation in the narrow angle cone. Furthermore, it is important to emphasize that the resonance escalation possesses a possibility to develop within two reaction channels. Thus, the first channel – delineates the occurrence that correlates to the spontaneous photon emission by an electron (laser-stimulated Compton interaction) with subsequent scattering on a nucleus (laser-assisted Mott procedure). The second channel – illustrates the configuration corresponding to the electron scattering on a nucleus with consecutive spontaneous photon emission in the wave field. Therefore, with implementation of the equivalent elementary criteria the value of the resonant frequency for the first channel perpetually represents a deteriorative counterpart to the second alternative. Moreover, the spontaneous photon radiation angle allocates a single-valued dependency with the resonant frequency for the first channel in contrast to the second displacement that categorizes a composite area with three various resonant frequency magnitudes for the particular emission angle diapason. The project data analysis proposes that the reaction channels do not interfere within the whole range of observation with a specific evaluation for the particles propagation at zero scattering angle. As a result of the investigation the calculations determine the scattering differential cross-section of the resonant construct development. To summarize, the particular cross-section within the resonant ambience significantly exceeds the according cross-section in the approximation of an external field absence. In conclusion, numerous scientific facilities with specialization in pulsed laser radiation (SLAC, FAIR, XFEL, ELI, XCELS) may experimentally validate the computational estimations.
\end{abstract}

\maketitle


\section{\label{sec:level1}Introduction}
An academic study of the Quantum Electrodynamic processes (QED) in the presence of a strong light field exemplifies an intensively advanced [6-44] scientific area and with implementation of the powerful laser radiation sources [1-5] in the contemporary applied and fundamental exploration fields indicates a possible prioritizing direction for the forthcoming elaborations. Various monographs [11-14] and reviews [15-21] systematize the principle results. The QED procedures of higher order with accordance to the fine structure constant in the field of an electromagnetic wave (laser-assisted QED effects) maintain a capacity to transmit within two potential options: resonant and non-resonant. In addition, the light field sustains a basis for the lower order with respect to the fine structure constant occurrences (laser-stimulated QED interactions) [15] and consequent appearances of the Oleinik resonances [9, 10]. It is important to emphasize that probability of the resonant transaction for the QED mechanism in the external field may substantially exceed by several orders of magnitude the correlative propositional scheme in the absence of a wave field [19, 20]. However, the preceding scrutiny of the Oleinik resonances appearances distinctively focused on a single channel of the RSB system when the electron radiates a spontaneous photon and sequentially scatters on a nucleus. In contrast, the second channel that interprets the case of electron scattering on a nucleus with subsequent spontaneous photon emission maintains a certain prospect for the physical contemplation. Furthermore, the explorations of the first channel of interaction characterized isolatedly not very high-energies of electrons with extensive scattering angles of the final electron [12-14, 19-20]. The recent work accentuates the essential demand in the investigation advancement [44].\\
The actual analysis develops the theory of the resonant RSB of ultrarelativistic electrons with considerable energetic magnitudes within the scope of scattering in the Coulomb field of a nucleus at small-scaled angles in the presence of a plane electromagnetic wave. Furthermore, the research scrutinizes both of the probable reaction channels. The investigation evaluates the arrangement in which the final electron and spontaneous photon propagate in the narrow angle cone coordinately with the momentum of the initial electron.\\
The study of the RSB of an electron on a nucleus in the external field utilizes a pair of characteristic parameters. Thus, the classical relativistically-invariant parameter [15, 19-22]:
\begin{equation} \label{eq:1}
\eta  = \frac{{eF\mathchar'26\mkern-10mu\lambda  }}{{m{c^2}}}
\end{equation}
functionally equal to the ratio of the work of the field at a wavelength to the rest energy of an electron (where $e$ and $m$ - are the charge and mass of an electron, $F$ and $\mathchar'26\mkern-10mu\lambda = c/\omega$ - correspond to the electric field strength and the wavelength, and $\omega$ - is the frequency of a wave). In addition, the research applies the coterminous quantum multiphoton parameter (Bunkin-Fedorov parameter) [8, 11, 19-22]:
\begin{equation} \label{eq:2}
\gamma_i = \eta \frac {m v_i c}{\hbar \omega}
\end{equation}
Where $v_i$  - is the velocity of the initial electron. Within the spectrum range of the optical frequencies ($\omega \sim 10^{15} s^{-1}$) the classical parameter is $\eta \sim 1$ for the fields of $F \sim 10^{10} \div 10^{11} V/cm$, and the quantum parameter is $\gamma_i \sim 1$ for the fields of $F \sim (10^5 \div 10^6)(c/v_i) V/cm$. Therefore, for the fields with $\eta \ll 1$ the quantum parameter  $\gamma_i$ may achieve a substantial magnitude order. However, the represented interpretation is applicable for the considerable angles of electron scattering on a nucleus exceptionally. The indicated exemplification allocates the Bunkin-Fedorov quantum parameter as the principal criterion defining the multiphoton effects. Accordingly, the research examines the phenomenon for the primary solution in the diapason of the moderately-strong field intensities that fulfills the specific prerequisites:
\begin{equation} \label{eq:3}
\eta \ll 1 , \gamma_i \gtrsim 1
\end{equation}
Notwithstanding, the quantum Bunkin-Fedorov parameter does not appear in the RSB process of electron scattering on a nucleus at small-scaled angles. The implied ambience classifies the appropriation of the classical parameter $\eta$ for the multiphoton effects characterization and conditions (3) establish an approximately weak external field.\\
The calculations in the article develop within the framework of the relativistic system of units: $\hbar=c=1$.

\section{\label{sec:level1}The amplitude of the RSB process of an electron on a nucleus in a light field}

The research selects the consequential form for the 4-potential of an external elliptically polarized light wave propagating complementary to the $z$ axis:
\begin{equation} \label{eq:4}
A({\phi})={\frac {F} {\omega}} {\cdot} (e_x \cos{\phi} + \delta e_y \sin{\phi}), \: \phi=kx={\omega(t-z)}
\end{equation}
Where $\delta$ - is the wave ellipticity parameter ($\delta = 0$ - determines the linear polarization, $\delta = \pm 1$ - delineates the circular polarization), $e_{x,y}=(0,{\bf e}_{x,y})$ and $k={\omega n}={\omega (1,{\bf n})}$ - designate the 4-vectors of polarization and the momentum of photon in the electromagnetic field, particularly: $k^2=0, \: e_{x,y}^2=-1, \: e_{x,y}k=0$.
The investigation analyzes the RSB kinematics within the Born approximation conception of electrons interaction with the field of a nucleus. Coordinately, a pair of Feynman diagrams illustrate (see Fig. 1) the compositional effect of the second order processes with accordance to the fine structure constant. The following expressions derivate the amplitude of the formalistic construct (see, for example, [19]):
\begin{equation} \label{eq:5}
S_{fi} = \sum_{n=-\infty}^{\infty} S_l
\end{equation}
where the subsequent equation exemplifies the partial amplitude with radiation and absorption of the $|l|$-photons of the wave:
\begin{equation} \label{eq:6}
S_l = -i \cdot {\frac {8 \pi^{5/2} \cdot Z e^3} {\sqrt{2 \omega' \tilde{E}_i \tilde{E}_f}}} \cdot e^{i \varphi_{fi}} \cdot [\overline{u}_f M_l u_i] \cdot {\frac {\delta (q_0)} {{\bf q}^3}}
\end{equation}
\begin{equation} \label{eq:7}
\begin{aligned}
M_l = \sum_{r=-\infty}^{\infty} [M_{r+l} (p_f, q_i) \cdot {\frac {\hat{q}_i + m_{*}} {q^2_i - m^2_{*}}} \cdot F_{-r}(q_i, p_i) + \\
+ F_{-r}(p_f, q_f) \cdot {\frac {\hat{q}_f + m_{*}} {q^2_f - m^2_{*}}} \cdot M_{r+l} (q_f, p_i)]
\end{aligned}
\end{equation}

\begin{figure}[h!]

     \begin{center}
     \includegraphics[width=8cm]{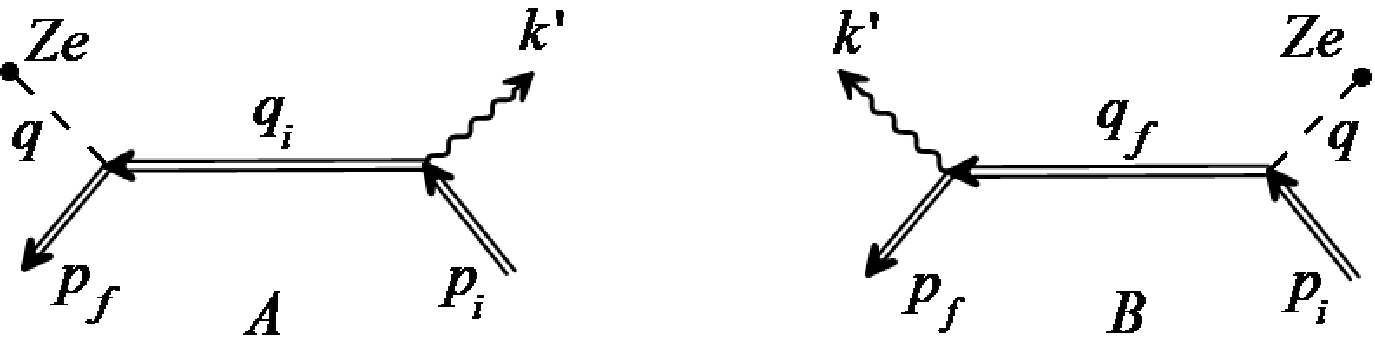}\label{Fig:Figure 1}
     \end{center}
     \caption{Feynman diagrams for the process of RSB of an electron on a nucleus in the field of a plane electromagnetic wave. The doublet outer lines characterize the Wolkow functions of an electron in the initial and final states, the intermediate line construes the Green function of an electron in the field of a plane wave (4). The waved lines illustrate the 4-momenta of the spontaneous photon and the dashed lines delineate the "pseudo-photon" of the nucleus recoil.}
  \end{figure}

In the equations (6)-(7) $\varphi_{fi}$ - is the independent from the summation indexes phase, $u_i, \overline{u}_f$ and $p_{i,f} = (E_{i,f}, {\bf p}_{i,f})$ - are the Dirac bispinors and the 4-momenta of the initial and final electrons. The sequential expressions determine the 4-momenta of the intermediate electrons $q_{i,f}$ and the transmitted 4-momentum $q$:
\begin{equation} \label{eq:8}
q_i = \tilde{p}_i - k' + rk,  \: q_f = \tilde{p}_f + k' - rk
\end{equation}
\begin{equation} \label{eq:9}
q = \tilde{p}_f - \tilde{p}_i + k' + lk
\end{equation}
Where $k' = \omega' (1,{\bf n}')$ - is the 4-momentum of the spontaneous photon, $\tilde{p}_{i,f}$ - are the 4-quasi-momenta, $m_{*}$ - is the effective mass of an electron in the field of a plane wave:
\begin{equation} \label{eq:10}
\tilde{p}_{i,f} = p_{i,f} + (1 + \delta^2) \eta^2 {\frac {m^2} {4(k p_{i,f})}} k
\end{equation}
\begin{equation} \label{eq:11}
\tilde{p}_{i,f} = m^2_{*}, \: m_{*} = m {\sqrt{1 + {\frac {1} {2}}(1 + \delta^2) \eta^2}}
\end{equation}
In the current work the mathematical expressions with a sign of a hat as in the ratio (7) indicate the scalar production of the 4-vector with Dirac gamma-matrices: $\tilde{\gamma}^{\mu} = (\tilde{\gamma}^{0}, \tilde{\bf \gamma})$, $\mu = 0, 1, 2, 3$. For example, $\hat{q}_i = q_{i \mu} \tilde{\gamma}^{\mu} = q_{i 0} \tilde{\gamma}^{0} - {\bf q}_i {\bf \tilde{\gamma}}$. The following statements detail the amplitudes $M_{r+l}$ and $F_r$ (see Fig. 1) from the equation (7):
\begin{equation} \label{eq:12}
\begin{aligned}
M_{r+l} (p_2, p_1) = a^0 \cdot L_{r+l}(p_2, p_1) + b_{-}^0 \cdot L_{r+l-1} + \\ + b_{+}^0 \cdot L_{r+l+1} + c^0 \cdot (L_{r+l-2} + L_{r+l+2})
\end{aligned}
\end{equation}
\begin{equation} \label{eq:13}
\begin{aligned}
F_{-r} (p_2, p_1) = (a \varepsilon^{*}) \cdot L_{-r}(p_2, p_1) + (b_{-} \varepsilon^{*}) \cdot L_{-r-1} + \\ + (b_{+} \varepsilon^{*}) \cdot L_{-r+1} + (c \varepsilon^{*}) \cdot (L_{-r-2} + L_{-r+2})
\end{aligned}
\end{equation}
Where $\varepsilon^{*}_{\mu}$ - is the 4-vector of polarization of spontaneous photon. The consequent proportions define the $a^{\mu}, b^{\mu}_{\pm}, c^{\mu}$ matrices
\begin{equation} \label{eq:14}
a^{\mu} = \tilde{\gamma}^{\mu} + (1+{\delta}^2) \cdot {\eta}^2 {\frac {m^2} {4(kp_1)(kp_2)}} k^{\mu}
\end{equation}
\begin{equation} \label{eq:15}
b^{\mu}_{\pm} = {\frac {1} {4}} \eta m \cdot [{\frac {\hat{\varepsilon}_{\pm} \hat{k} \tilde{\gamma}^{\mu}} {(kp_2)}} + {\frac {\tilde{\gamma}^{\mu} \hat{k}  \hat{\varepsilon}_{\pm} } {(kp_1)}}], \: \hat{\varepsilon}_{\pm} = \hat{e}_{x} \pm i \delta \cdot \hat{e}_{y}
\end{equation}
\begin{equation} \label{eq:16}
c^{\mu} = - (1 -{\delta}^2) \cdot {\eta}^2 {\frac {m^2} {8(kp_1)(kp_2)}} \cdot k^{\mu}
\end{equation}
The subsequent equations characterize the special functions $L_{r+l}$ and $L_{-r}$ and their arguments [47]:
\begin{equation} \label{eq:17}
L_{r'}(p_2,p_1) = e^{-i r' \chi} \cdot \sum_{S=-\infty}^{\infty} e^{2is\chi} \cdot J_{r'-2s}(\gamma) \cdot J_s(\beta)
\end{equation}
\begin{equation} \label{eq:18}
tg \chi = \delta \cdot {\frac {(e_y Q)} {(e_x Q)}}, \: Q = {\frac {p_2} {(kp_2)}} - {\frac {p_1} {(kp_1)}}
\end{equation}
\begin{equation} \label{eq:19}
\gamma = \eta m \sqrt{(e_x Q)^2 + \delta^2 \cdot (e_y Q)^2}
\end{equation}
\begin{equation} \label{eq:20}
\beta = {\frac {1} {8}} (1-\delta^2) \eta^2 m^2 \cdot [{\frac {1} {(kp_2)}} - {\frac {1} {(kp_1)}}]
\end{equation}
Additionally, the amplitudes $M_{r+l}(p_f,q_i)$ and $M_{r+l}(q_f,p_i)$ in the ratios (12), (14)-(20) designate the rearrangements $p_1 \rightarrow q_i$, $p_2 \rightarrow p_f$ and $p_1 \rightarrow p_i$, $p_2 \rightarrow q_f$, and the amplitudes $F_r(q_i,p_i)$ and $F_r(p_f,q_f)$ in the statements (13), (14)-(20) indicate the substitutions  $p_1 \rightarrow p_i$, $p_2 \rightarrow q_i$ and $p_1 \rightarrow q_f$, $p_2 \rightarrow p_f$.

\section{\label{sec:level1}The poles of the RSB amplitude}

The quasi-discrete construct of the system delineates the resonant organization of the amplitude (3.8)-(3.9): electron + plane electromagnetic wave. As a result, the 4-momentum of the intermediate electron emerges to the mass shell surface with the condition of the energy-momentum conservation laws fulfillment.\\
The consequential investigation concentrates on the resonances in the area of the laser fields (3). Therefore, the specified ambience consents the neglection in the 4-quasi-momenta (10) of the addends that are proportional to the parameter $\eta^2 \ll 1$. Particularly, within the resonance the following requirements (see Fig. 2) realize for the first and second terms in the amplitude (6)-(7):
\begin{equation} \label{eq:21}
q_i^2 = m^2
\end{equation}
\begin{equation} \label{eq:22}
q_f^2 = m^2
\end{equation}

 \begin{figure}[h!]
       \begin{center}
     \includegraphics[width=8cm]{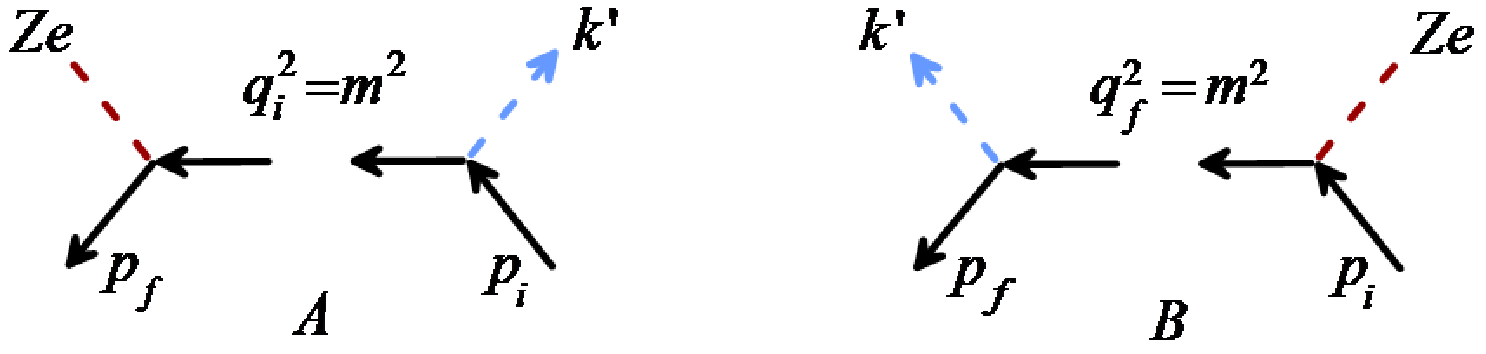}\label{Fig:Figure 2}
     \end{center}
     \caption{Resonant spontaneous bremsstrahlung of an electron in the field of a nucleus and a plane electromagnetic wave.}
  \end{figure}

\noindent The preceding investigations scrutinized the resonances (21) for the first amplitude addend (6)-(7) (see Fig. 2, diagram A) that characterize the process of spontaneous photon radiation by an electron with subsequent electron scattering on a nucleus at considerable angles for the various electron energies with exclusion of the instance of photon emission at narrow angle cone in the direction of the initial ultrarelativistic electron propagation $E_i \gtrsim m^2/\omega \sim 10^5 \div 10^6 MeV$. Contrastingly, the resonances (22) for the second term of the amplitude (6)-(7) (see Fig. 2, diagram B) that construe the occurrence of electron scattering on an nucleus with sequential spontaneous photon emission represent a certain prospect for the actual scientific endeavour. The study derives the expressions that determine the 4-momenta of the intermediate electrons $q_i$ and $q_f$ (8) and the transferred 4-momentum $q$ (9) for the diagrams A and B (see Fig. 2) in the following form:
\begin{equation} \label{eq:23}
p_i + rk = q_i + k{'}
\end{equation}
\begin{equation} \label{eq:24}
q = p_f + q_i + (l+r)k
\end{equation}
and
\begin{equation} \label{eq:25}
q_f + rk = p_f+k{'}
\end{equation}
\begin{equation} \label{eq:26}
q = q_f - p_i + (r+l)k
\end{equation}
The equations (23) and (25) are relevant with singular limitation $r \geqslant 1$ correlating to the prerequisite application $p_i^2 = q_i^2 = m^2$, $p_f^2 = q_f^2 = m^2$ and $k^2 = k'^2 = 0$.
The represented regulations and the structure of the amplitude (6)-(7) indicate (see, additionally, Fig. 2) that the magnitude $F_r$ (13) with accordance to the fulfillment of the 4-momentum conservation laws (23) or (25) in the resonance delineates the amplitude of the process of the photon radiation with the 4-momentum $k'$ by an electron as a result of the absorption of r -photons of the wave. Various scientists (Nikishov, Ritus and etc.) examined the illustrated effect (see review [15]). The quantity of the $M_{r+l}$ (12) physical value and the transmitted 4-momentum $q$ (24) or (26) define the amplitude of the electron scattering on a nucleus in the external field with emission and absorption of $|r+l|$ wave photons.  The phenomenon underwent a precise examination in the nonrelativistic case by Bunkin and Fedorov [8] and for the general relativistic case by Denisov and Fedorov [11]. Consequently, with exclusion of the amplitudes A and B interference at zero angle of emission (see Fig. 2) the process of the RSB of an electron on a nucleus in the light field effectively splits into two subsequential first-order processes with accordance to the fine structure constant: electron scattering on a nucleus and radiation of a photon with 4-momentum $k'$ by an electron in the field of a wave (see Fig. 2). In addition, for the arrangement of coincidence of the propagation directions of the spontaneous and external field photons the simultaneous fulfillment of the resonant conditions (21) or (22) and the 4-momentum conservation laws (23) or (25) is inaccessible. Therefore, the resonances occur exceptionally for the non-parallel motion of the registered photons. For the specified status of the contemplated development of spontaneous photon radiation by an electron in the laser field the classical parameter $\eta$ exemplifies the multiphoton parameter properties. Thus, for the fields (3) electron absorbs primarily $r = 1$ amount of photons from the external wave. Comparatively, for the process of electron scattering at considerable angles the quantum Bunkin-Fedorov parameter $\gamma_i \gtrsim 1$ complies equally to the multiphoton parameter. As a result, an electron possesses a possibility to absorb and radiate a substantial quantity of wave photons. However, for the situation of electron scattering at moderate angels the classical parameter $\eta$ constitutes the multiphoton parameter functions and with application of the conditions (3) the maximally probable process comprehends a single light field photon. \\
The research examines the expressions (21), (23) and (22), (25) for the derivation of the spontaneous photon frequency in resonance for the diagrams A and B (see Fig. 2):
\begin{equation} \label{eq:27}
\omega'_{(a)} = r \omega {\frac {\kappa_i} {\kappa'_i + 2r\omega \sin^2 (\theta'/2)}}
\end{equation}
\begin{equation} \label{eq:28}
\omega'_{(b)} = r \omega {\frac {\kappa_f (\omega')} {\kappa'_f(\omega') - 2r\omega \sin^2 (\theta'/2)}}
\end{equation}
Where:
\begin{equation} \label{eq:29}
\kappa_i = E_i - |{\bf p}_i| \cos \theta_i, \: \kappa'_i = E_i - |{\bf p}_i| \cos \theta'_i,
\end{equation}
\begin{equation} \label{eq:30}
\kappa_f (\omega') = E_f - |{\bf p}_f| \cos \theta_f, \: \kappa'_f (\omega') = E_f - |{\bf p}_f| \cos \theta'_f,
\end{equation}
\begin{equation} \label{eq:31}
\theta_j = \measuredangle ({\bf k},{\bf p}_j), \: \theta'_j= \measuredangle ({\bf k'},{\bf p}_j), \: j = i, f; \: \theta'= \measuredangle ({\bf k'},{\bf k})
\end{equation}
For the channel A the initial electron energy and the collateral angles of scattering of the initial electron and spontaneous photon definitely establish the resonant frequency (27). Previous works evaluate the effects in details [12-14, 19]. Contrastingly, for the channel B the spontaneous photon frequency (28) is a composite function. Accordingly, the energy and momentum of the final electron depend on the frequency of the spontaneous photon as a materialization of the energy conservation law for the provided effect.
\begin{equation} \label{eq:32}
E_f = E_i - \omega' + l\omega \approx E_i -\omega'
\end{equation}
Thus, the right part of the expression (28) depends on the spontaneous photon frequency. In addition, the calculations propose that the accounting of the quantity of the emitted and absorbed photons of the wave may be neglected from the examination of the equation (32) $l\omega / E_i \lesssim \gamma_i \omega / E_i \approx \eta m / E_i \ll 1$. Within the range of the optical frequencies the second addend in the denominator of the expression (28) delineates a substantially minor magnitude in comparison to the first term for the non-relativistic and relativistic electron energies. Consequently, for the selection of non-relativistic electron ambience the statement (28) realizes the configuration:
\begin{equation} \label{eq:33}
\omega'_{(b)} \approx r \omega
\end{equation}
And for the relativistic electron energies the (28) reconstructs into:
\begin{equation} \label{eq:34}
\omega'_{(b)} \approx r \omega {\frac {(1 - v_f \cos \theta_f)} {(1 - v_f \cos \theta'_f)}} \sim r \omega
\end{equation}
Where $v_f = |{\bf p}_f|/E_f$. Therefore, the resonance for non-relativistic electrons appears at the frequency interval that is directly proportional to the frequency of a laser wave, and for relativistic electrons the resonant status emanates from the frequency diapason comparable to the ratio order of the laser frequency multiplier.\\
The subsequent analysis scrutinizes the prominent interaction with maximal experimental relevance for ultrarelativistic electrons distribution when the spontaneous photon and final electron propagate within a narrow cone in accordance to the initial electron momentum vector direction.
\begin{equation} \label{eq:35}
E_{i,f} \gg m
\end{equation}
\begin{equation} \label{eq:36}
\theta'_{i,f} \ll 1, \: \theta = \measuredangle ({\bf p}_i,{\bf p}_f) \ll 1, \: \theta' \sim 1
\end{equation}
The conditions (35), (36) constitute the basis for statements:
\begin{equation} \label{eq:37}
\kappa_i = 2 E_i \sin^2 {\frac {\theta_i} {2}}, \: \kappa'_i \approx {\frac {m^2} {2 E_i}} (1 + {\delta'_i}^2)
\end{equation}
\begin{equation} \label{eq:38}
\begin{aligned}
\kappa_f = 2 E_i (1 - {\frac {\omega'} {E_i}}) \sin^2 {\frac {\theta_f} {2}}, \: \kappa'_f \approx {\frac {m^2} {2 E_i (1 - \omega'/E_i)}} \cdot \\ \cdot [1 + {\delta'_f}^2(1 - {\frac {\omega'} {E_i}})^2]
\end{aligned}
\end{equation}
Where
\begin{equation} \label{eq:39}
\delta'_i = {\frac {E_i \theta'_i} {m}}, \: \delta'_f = {\frac {E_i \theta'_f} {m}}
\end{equation}
With reconsideration of the (27), (37) equations the calculations derive the resonant frequency of the spontaneous photon for the channel A (see Fig. 2):
\begin{equation} \label{eq:40}
x'_{(a)}(\delta{'}^2_i)= {\frac {\varepsilon_i} {1+\varepsilon_i+\delta{'}^2_i}}, \: x'_{(a)}={\frac {\omega'_{(a)}} {E_i}}
\end{equation}
\begin{equation} \label{eq:41}
\varepsilon_i = {\frac {E_i} {E_{*}}}, \: E_{*} = {\frac {m^2} {4 \omega \sin^2(\theta{'}/2)}}
\end{equation}
Where $\varepsilon_i$ - is the characteristic parameter equal to the ratio of the initial electron energy to the characteristic energy of the process $E_{*}$. The rest energy of an electron, the frequency of a wave and the emission angle of the spontaneous photon with interdependency to the direction of the wave propagation define the represented energy. In the optical frequencies range the characteristic energy of the phenomenon obtains the order of  $E_{*} \sim 10^5 \div 10^6 MeV$. The authors of the articles [12-14, 19] inspected the occurrence when: $\varepsilon_i \ll 1 \: (E_i \ll E_{*})$ designating the moderate initial energies of ultrarelativistic electrons and the resonant frequency of the spontaneous photon was equal to: $\omega'_{(a)} \sim \varepsilon_i E_i \ll E_i$. Moreover, in the previous investigations the electron scattered at considerable angles.\\
The actual research examines the effect for: $\varepsilon_i \gtrsim 1 \: (E_i \gtrsim E_{*})$ construing the conception with considerable energies of the initial electrons and small-scaled scattering angles. The expression (40) establishes the single-valued functional proportionality for the spontaneous photon frequency to its angle of radiation that is relative to the initial electron momentum. Thus, for the arrangement when the spontaneous photon emerges in the coincidental coordinate orientation as the initial electron momentum $(\delta{'}^2_i = 0)$ the resonant frequency attains the maximal magnitude:
\begin{equation} \label{eq:42}
x'_{(a)}(0) = x'_{max}={\frac {\varepsilon_i}  {1+\varepsilon_i}}
\end{equation}
Increase of the radiation angle of the spontaneous photon produces the contraction of the resonant frequency (40) and with  $\delta{'}^2_i \rightarrow  \infty$ the magnitude tends to zero.\\
The revision of the statements (28), (32), (38) for the resonant frequency of the channel B formulates the cubic equation that acquires the configuration:
\begin{equation} \label{eq:43}
\delta{'}^2_f x{'}^3_{(b)} - 2\delta{'}^2_f x{'}^2_{(b)} + (1 + \delta{'}^2_f + \varepsilon_i) x'_{(b)} - \varepsilon_i = 0, \:  x'_{(b)} = {\frac {\omega'_{(b)}} {E_i}}
\end{equation}
The research analyzes the equation (43) for the real roots within the interval $0<x'_{(b)res}<1$.\\
The subsequent derivations indicate that the solutions of (43) essentially depend of the magnitudes of the parameters $\varepsilon_i$ and $\delta{'}^2_f$. However, the characteristic parameter   delineates the resonant frequency and the angle of emission of the spontaneous photon. The evaluation specifies that the equation (43) retrieves a single real root for the parameter value: $\delta{'}^2_f = 0$ illustrating the instance of the spontaneous photon propagation in parallel to the momentum of the final electron. The accentuated conditions proffer the maximal magnitude for the resonant frequency of the spontaneous photon in the channel B $x'_{(b)} = x'_{max}$ that coincides with the obtained value for the channel A (see (42)).
The consequent examination of the equation (43) utilizes the computations with $\delta{'}^2_f \neq 0$. Thus, the functional limitation authorizes the division of the expression (43) addends on $\delta{'}^2_f$. Additionally, the research arranges the substitution of the variable.
\begin{equation} \label{eq:44}
x'_{(b)} = y + {\frac {2} {3}}
\end{equation}
The calculations disposition the following statement:
\begin{equation} \label{eq:45}
y^3 + ay + b = 0
\end{equation}
\begin{equation} \label{eq:46}
a = {\frac {1} {3 \delta{'}^2_f}} [3(1+\varepsilon_i)-\delta{'}^2_f], \: b = {\frac {1} {27 \delta{'}^2_f}} [2(9+\delta{'}^2_f)-9\varepsilon_i]
\end{equation}
the solutions of the (45) with accordance to the (44) are:
\begin{equation} \label{eq:47}
x'_{(b)j} = y_j + {\frac {2} {3}}, \: j = 1, 2, 3
\end{equation}
\begin{equation} \label{eq:48}
y_1 = \alpha_{+} + \alpha_{-}, \: y_{2,3} = -{\frac {1} {2}} (\alpha_{+} + \alpha_{-}) \pm i{\frac {\sqrt{3}} {2}}(\alpha_{+} - \alpha_{-})
\end{equation}
\begin{equation} \label{eq:49}
\alpha_{\pm} =  [-{\frac {b} {2}} \pm \sqrt{Q}]^{1/3}, \: Q = ({\frac {a} {3}})^3 + ({\frac {b} {2}})^2
\end{equation}
The area of the determinant dispersion $Q \geqslant 0$ indicates a single solution and $Q < 0$ produces three various roots. The investigation precisely explores the result system. The pattern spectrum:
\begin{equation} \label{eq:50}
Q < 0
\end{equation}
realizes when the parameter $a < 0$. Therefore:
\begin{equation} \label{eq:51}
\delta{'}^2_f = 3(1+\varepsilon_i)+\overline{d}, \: \overline{d} > 0
\end{equation}
The consequential derivations estimate the displacement of (51) into the expression (50) and accumulation of the square inequality for the parameter $\overline{d}$:
\begin{equation} \label{eq:52}
(\overline{d} - \overline{d}_{-})(\overline{d} - \overline{d}_{+})<0
\end{equation}
Where
\begin{equation} \label{eq:53}
\overline{d}_{\pm} = {\frac {(\varepsilon_i - 8)} {8}} [(\varepsilon_i + 4) \pm \sqrt{\varepsilon_i(\varepsilon_i - 8)}]
\end{equation}
The evaluation establishes that $\varepsilon_i>8$ and $\overline{d}_{-}<\overline{d}<\overline{d}_{+}$. Furthermore, with reconsideration of the ratios (47)-(49) the research procures three probable propositions for the resonant frequency of the spontaneous photon:
\begin{equation} \label{eq:54}
\begin{aligned}
x'_{(b)1} = {\frac {2} {3}} + d' \cos({\frac {\varphi'} {3}}), \: x'_{(b)2} = {\frac {2} {3}} + d' \cos({\frac {\varphi'} {3}} +{\frac {2\pi} {3}}), \: \\ x'_{(b)3} = {\frac {2} {3}} + d' \cos({\frac {\varphi'} {3}} +{\frac {4\pi} {3}})
\end{aligned}
\end{equation}
\begin{equation} \label{eq:55}
\begin{aligned}
d' = {\frac {2} {3 \delta'_f}} \sqrt{\delta{'}^2_f - 3(1 + \varepsilon_i)}, \: \cos \varphi' = {\frac {\delta'_f [9 \varepsilon_i - 2(9 + \delta{'}^2_f)]} {2[\delta{'}^2_f -3(1 + \varepsilon_i)]^{3/2}}}, \: \\ 0\leq \varphi' \leq \pi
\end{aligned}
\end{equation}
Where the resonant angle of the spontaneous photon radiation that defines the resonant frequencies (54) in accordance to the momentum of the final electron locates within the diapason:
\begin{equation} \label{eq:56}
\delta{'}^2_{-} < \delta{'}^2_f < \delta{'}^2_{+}
\end{equation}
\begin{equation} \label{eq:57}
\delta{'}^2_{\pm} = 3 ( 1 + \varepsilon_i ) + {\frac {1} {8}} (\varepsilon_i - 8) [(\varepsilon_i + 4) \pm \sqrt{\varepsilon_i (\varepsilon_i - 8)}], \: \varepsilon_i > 8
\end{equation}
Subsequently, the analysis scrutinizes the area of:
\begin{equation} \label{eq:58}
Q \geqslant 0
\end{equation}
The indicated framework maintains a single solution for the resonant frequency of the spontaneous photon:
\begin{equation} \label{eq:59}
x'_{(b)} =  {\frac {2} {3}} + (\alpha_+ + \alpha_-)
\end{equation}
The equations (46), (49) specify the $\alpha_{\pm}$. Additionally, distribution of the possible spontaneous photon angles of emission $(\delta{'}^2_f)$ substantially depends on the magnitude of the parameter $\varepsilon_i$:
\begin{equation} \label{eq:60}
0 < \delta{'}^2_f \leq 3(1 + \varepsilon_i), \: if \: 0 < \varepsilon_i \leq 8
\end{equation}
\begin{equation} \label{eq:61}
0 < \delta{'}^2_f \leq \delta{'}^2_{-}, \: \delta{'}^2_{+} \leq \delta{'}^2_f < \infty, \: if \: \varepsilon_i > 8
\end{equation}
The Fig. 3 illustrates the spectrum of the potential resonant radiation angles of the spontaneous photon in connection to the characteristic parameter $\varepsilon_i$ (41) that approximates whether a single $(Q > 0)$ or three $(Q < 0)$ various value alternatives of the resonant frequency emanate from the experimental ambience.\\
The study indicates two sectors of characteristic parameter allocation with dispersion structures that differentiate qualitatively for the resonant frequencies and the spontaneous photon emission angles. Consequently, within the range of $\varepsilon_i \leqslant 8$ (60) the radiation angle with interdependency to the final electron momentum definitely designates the spontaneous photon resonant frequency. Therefore, within the highlighted limitations the spontaneous photon resonant frequency fluctuates from the maximum $x'^{max}_{res}$ (42) when $\delta'_f = 0$ to the minimum:
\begin{equation} \label{eq:62}
x'_{(b)min} = {\frac {1} {3}} [2 - ({\frac {8 - \varepsilon_i} {1 + \varepsilon_i}})^{\frac {1} {3}}]
\end{equation}
for the maximal degree of the emission angle the parameter is $\delta{'}^2_f = \delta{'}^2_{fmax} = 3(1+\varepsilon_i)$. For the radiation angles of $\delta{'}^2_f > \delta{'}^2_{fmax}$ the resonant state in the channel B is not achievable (see Fig. 4). \\

Within the area of the characteristic parameter magnitude $\varepsilon_i > 8$ the attitude of the spontaneous photon resonant frequency as a function of the emission angle in accordance to the final electron momentum varies significantly (see Fig. 5). Thus, for the radiation angles from the interval:
\begin{equation} \label{eq:63}
0 \leq \delta'^2_f < \delta'^2_{-}
\end{equation}
The radiation angle accurately specifies the resonant frequency of the spontaneous photon and the magnitude modulates from the maximal $x'_{max}$ to $x'_{(b)-}$. For the angle $\delta{'}^2_f = \delta{'}^2_{-}$ the resonance frequency acquires two available values. Within the range of the emission angles:
\begin{equation} \label{eq:64}
\delta{'}^2_{-} < \delta{'}^2_f < \delta{'}^2_{+}
\end{equation}
the resonant frequency modifies from $x'_{(b)-}$ to $x'_{(b)+}$ obtaining three computational alternatives for every parameter $\delta{'}^2_f$ formation. For the radiation angle $\delta{'}^2_f = \delta{'}^2_{+}$ the resonant frequency produces two results. Finally, for the diapason:
\begin{equation} \label{eq:65}
\delta{'}^2_{f} > \delta{'}^2_{+}
\end{equation}
similarly to the (63) composition the radiation angle delineates the resonant frequency gradually decreasing from $x'_{(b)+}$ to $x'_{(b)} \ll 1$. The resonant frequencies $x'_{(b) \pm}$ arrange from the substitution $\delta{'}^2_f = \delta{'}^2_{\pm}$ into solutions (49), (59). \\
Precise examination of the resonant frequencies for the channels A and B (see Fig. 4 and Fig. 5) demonstrates that the frequencies coincide exceptionally when the spontaneous photon propagates in the coequal direction to the momenta of the initial and final electrons. The indicated effect physically conceptualizes the scattering of an electron on a nucleus at zero emission angle $(\delta{'}^2_f = \delta{'}^2_{i} = 0)$ and radiation of the spontaneous photon at maximal frequency (42). Therefore, with exclusion of the represented occurrence from the consideration the channels A and B do not interfere. Subsequent research concentrates on the exemplified mechanics $(\delta'^2_f \neq 0, \: \delta'^2_i \neq 0)$.

  \begin{figure}[h!]
       \begin{center}
     \includegraphics[width=7cm]{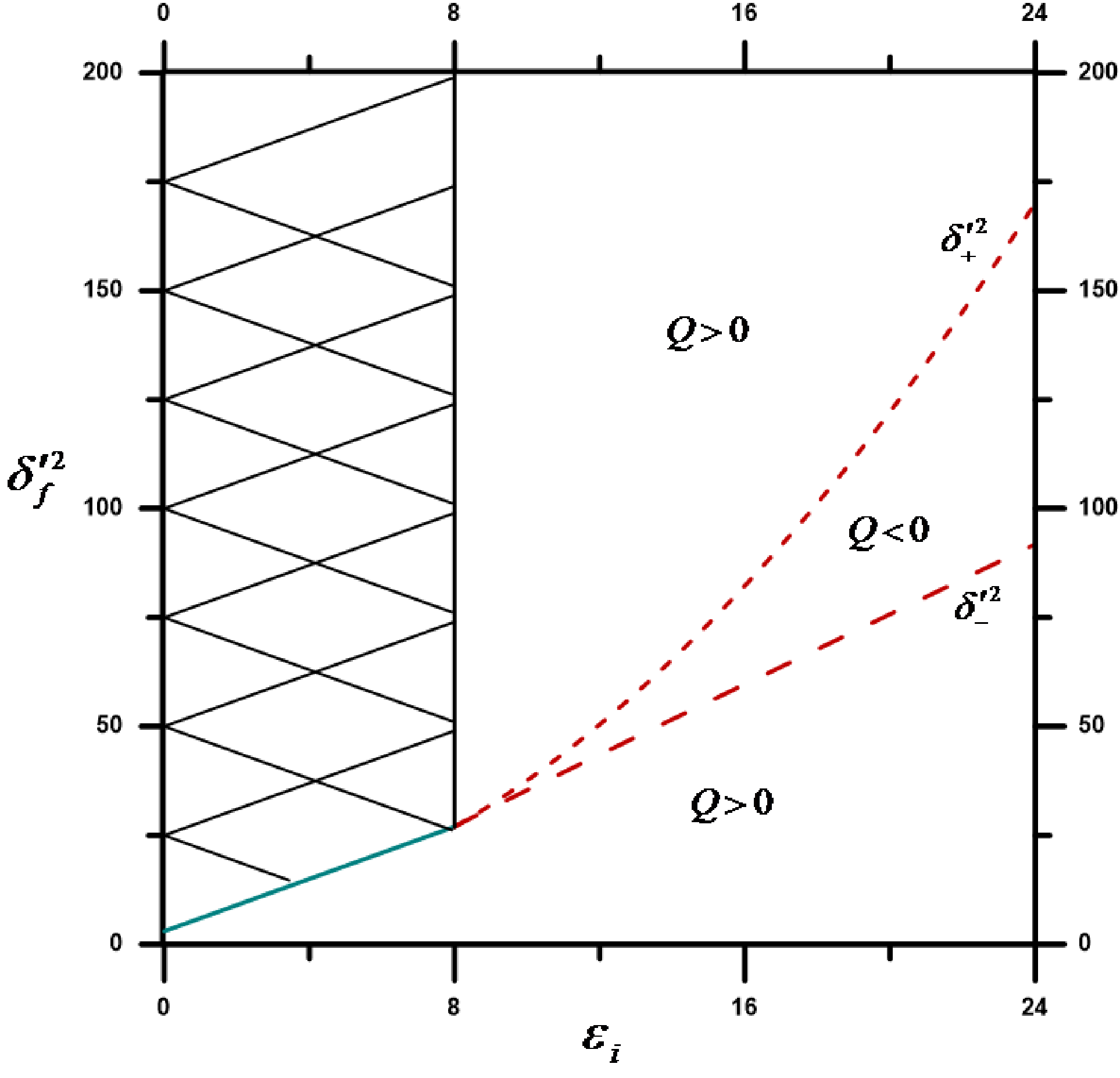}\label{Fig:Figure 3}
     \end{center}
     \caption{Dependency of the angle of emission of the resonant spontaneous photon (parameter $\delta'^2_f$) (56), (60) on the parameter $\varepsilon_i$ (43) for the complete spectrum of spontaneous photon resonant frequency magnitude. The diapason of parameters $\delta'^2_f$ and $\varepsilon_i$ with determinant section $Q \geqslant 0$ delineates a single possible resolution for the resonant frequency (59). The range of the parameters $\delta'^2_f$ and $\varepsilon_i$ with determinant value $Q < 0$ designates three various solutions of the resonant spontaneous photon frequency (54), (55). The effect of electron emission is absent for the shaded region.}
  \end{figure}

\begin{figure}[h!]
   \begin{minipage}{.5\textwidth}
       \includegraphics[width=7cm]{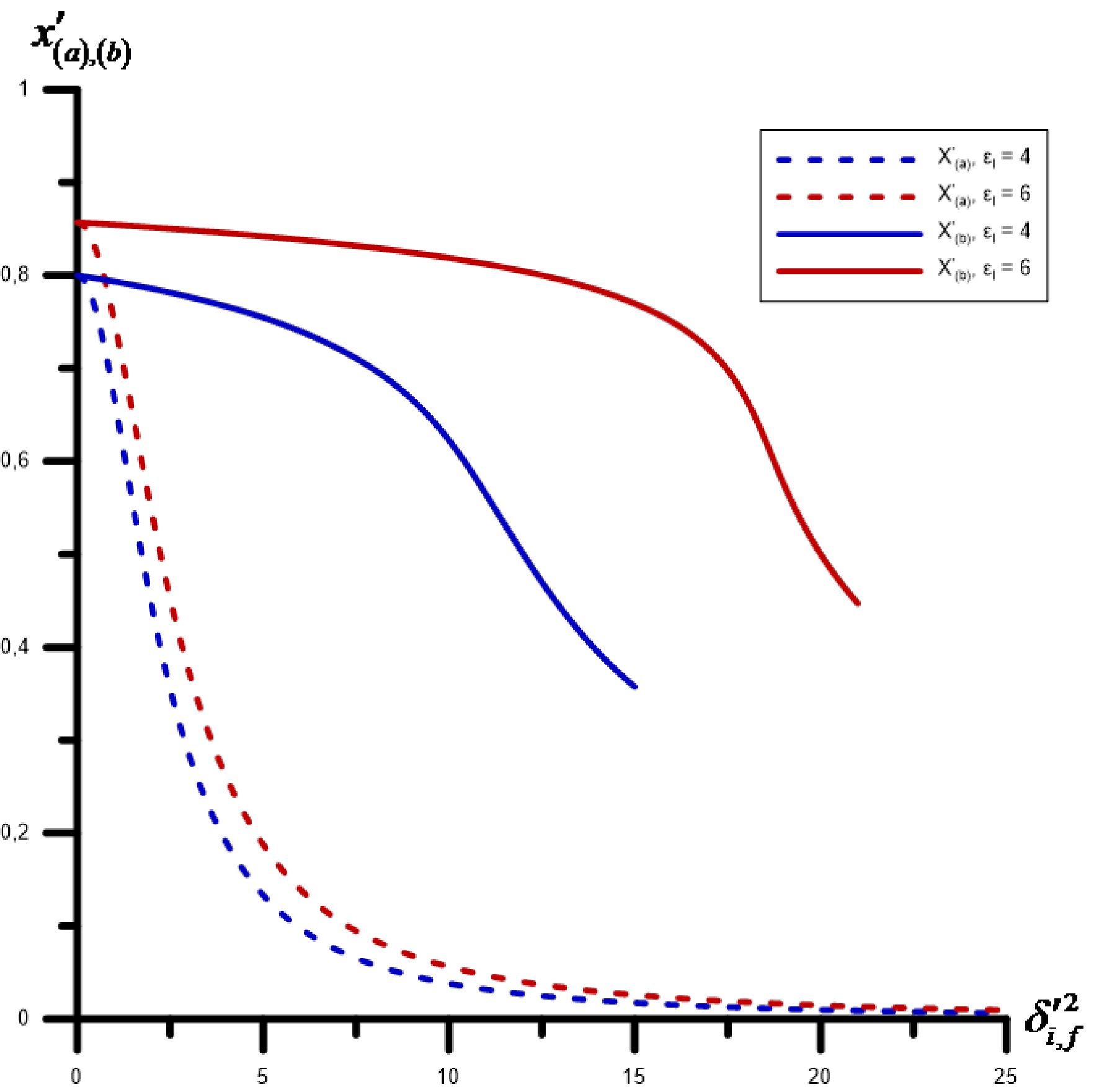}
  \end{minipage}
  \begin{minipage}{.5\textwidth}
      \includegraphics[width=7cm]{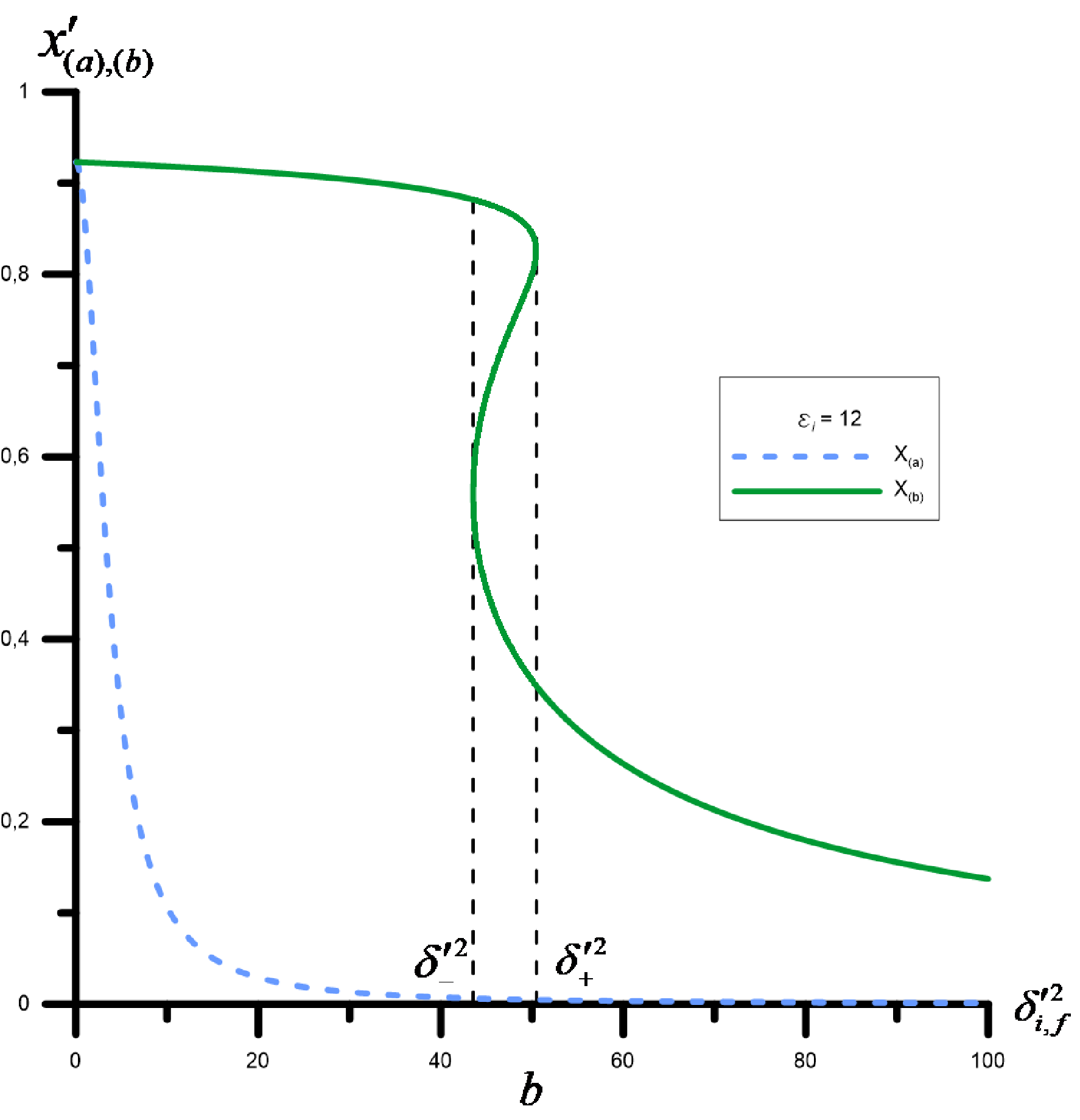}
  \end{minipage}
  \caption{The dependency of the spontaneous photon resonant frequency on the radiation angle. Fig. 4a - for the range of $\varepsilon_i \leq 8$ parameter magnitudes. Fig. 4b - for the continuum of $\varepsilon_i > 8 \: (\varepsilon_i = 12)$ parameter progression. The dashed lines characterize the resonant frequencies of the channel A $(x'_{(a)}(\delta{'}^2_i))$ (see (40)), the solid lines illustrate the resonant frequencies of the channel B  $(x'_{(b)}(\delta{'}^2_f))$ (see (54), (56) and (59)-(61)).}
  \label{Fig:Figure 4}
\end{figure}

\section{\label{sec:level1}The resonant differential scattering cross-section}

The consequential development of investigation arranges a simplification of the expression for the amplitude of the RSB process of scattering of ultrarelativistic electron (35), (36) on a nucleus in the field of a plane electromagnetic wave (3). The research derives following equation for the partial amplitude (6), (7), (12), (13) when $r = 1$:
\begin{equation} \label{eq:66}
S_l=-{i} {\cdot} {\frac {8 {\pi^{5/2}} \cdot Z e^3} {\sqrt{2 \omega' E_i E_f}}} {\cdot} e^{(i \varphi_{f,i})} {\cdot} {[\overline{u}_f M_l u_i]} {\frac {\delta (q_0)} {{\bf q}^2}}
\end{equation}
with accordance to the (66) statement the study acquires the proportion for the probability in a unit of time and in a single component of a volume by utilization of the standard procedure (see, for example, [19]). Subsequently, the calculations accumulate:
\begin{equation} \label{eq:67}
dw = \sum_{l=-\infty}^{\infty} dw_l
\end{equation}
Where the partial probability segment of composition is:
\begin{equation} \label{eq:68}
dw_l = {\frac {2 {\pi^{5/2}} (Z e^3)^2} {2 \omega' E_i E_f}} {\frac {|\overline{u}_f M_l u_i|^2} {{\bf q}^4}} \delta (q_0) d^3 p_f d^3 k'
\end{equation}
Here the particular formulation defines the matrix $M_l$:
\begin{equation} \label{eq:69}
M_l = {\frac {L_{1+l}(p_f, q_i)} {(q_i^2 - m^2)}} (\varepsilon_{\mu}^{*} G_i^{\mu}) + {\frac {L_{1+l}(q_f, p_i)} {(q_i^2 - m^2)}} (\varepsilon_{\mu}^{*} G_f^{\mu})
\end{equation}
\begin{equation} \label{eq:70}
G_i^{\mu} =  \tilde{\gamma}^{0} (\hat{q}_i + m) Y_i^{\mu}, \: G_f^{\mu} = Y_f^{\mu} (\hat{q}_f + m) \tilde{\gamma}^{0}
\end{equation}
The matrices $Y_i^{\mu}$ and $Y_f^{\mu}$ are:
\begin{equation} \label{eq:71}
\begin{aligned}
Y_i^{\mu} = \tilde{\gamma}^{\mu} \cdot a_0(q_i,p_i) + a_1(q_i,p_i) \cdot (k^{\mu} \hat{\varepsilon}_{+} - \varepsilon_{+}^{\mu} \hat{k}) + \\ + a_2(q_i,p_i) \cdot \hat{\varepsilon}_{+} \hat{k} \tilde{\gamma}^{\mu}
\end{aligned}
\end{equation}
\begin{equation} \label{eq:72}
\begin{aligned}
a_0 = L_{-1}(q_i,p_i), \: a_1 = \eta {\frac {m} {2(k p_i)}} L_0(q_i,p_i), \\ a_2 = {\frac {1} {4}} \eta m \cdot [{\frac {1} {2(k q_i)}} - {\frac {1} {2(k p_i)}}] \cdot L_0(q_i,p_i)
\end{aligned}
\end{equation}
and
\begin{equation} \label{eq:73}
\begin{aligned}
Y_f^{\mu} = \tilde{\gamma}^{\mu} \cdot a_0(p_f,q_f) + a_1(p_f,q_f) \cdot (k^{\mu} \hat{\varepsilon}_{+} - \varepsilon_{+}^{\mu} \hat{k}) + \\ + a_2(p_f,q_f) \cdot \hat{\varepsilon}_{+} \hat{k} \tilde{\gamma}^{\mu}
\end{aligned}
\end{equation}
\begin{equation} \label{eq:74}
\begin{aligned}
a_0 = L_{-1}(p_f,q_f), \: a_1 = \eta {\frac {m} {2(k q_f)}} L_0(p_f,q_f), \\ a_2 = {\frac {1} {4}} \eta m \cdot [{\frac {1} {2(k p_f)}} - {\frac {1} {2(k q_f)}}] \cdot L_0(p_f,q_f)
\end{aligned}
\end{equation}
Thus, the article analysis intensifies scientific endeavor in order to attain the probability of the resonant process without the amplitude interference (the first and second addends in (69)). The resonant probability for the channel A is:
\begin{equation} \label{eq:75}
\begin{aligned}
dw_{l(a)} = {\frac {2 \pi (Z e^3)^2} {2 \omega' E_i E_f}} {\frac {|L_{1+l}(p_f,q_i)|^2} {|q_i^2 - m^2|^2 {\bf q}^4}} \cdot \\ \cdot |\overline{u}_f (\varepsilon_{\mu}^{*} G_i^{\mu}) u_i|^2 \delta (q_0) d^3 p_f d^3 k'
\end{aligned}
\end{equation}
The paper framework limitation appropriates the non-polarized particles organization. Therefore, the computation of average polarization of initial electrons with summation of polarizations of final electrons and spontaneous photons comprehends a replacement:
\begin{equation} \label{eq:76}
\begin{aligned}
|\overline{u}_f (\varepsilon_{\mu}^{*} G_i^{\mu}) u_i|^2 = \varepsilon_{\mu}^{*} \varepsilon_v (\overline{u}_f G_i^{\mu} u_i)(\overline{u}_i \overline{G}_i^{v} u_f) \rightarrow \\ \rightarrow - {\frac {1} {2}} Sp[G_{i \mu}(\hat{p}_i + m)\overline{G}_i^{\mu}(\hat{p}_f + m)]
\end{aligned}
\end{equation}
The subsequential to the calculation of the spur of matrix (76) methods are the division of the expression (75) on the flux density of the incident particles $v_i = |{\bf p}_i| / E_i$ and integration on the energy of the final electron with application of the Dirac delta function $\delta (q_0)$. As a result, the resonant differential cross-section for the channels A $(d\sigma_{(a)res})$ and B $(d\sigma_{(b)res})$ of the RSB phenomenon obtains a form:
\begin{equation} \label{eq:77}
d\sigma_{res} =  d\sigma_{(j)res}, \: j = a,b
\end{equation}
Where
\begin{equation} \label{eq:78}
d\sigma_{(a)res} = d\sigma (p_f,q_i) \cdot {\frac {2 \pi m |{\bf q}_i|} {|q_i^2 - m^2|^2}} \cdot dW'_1 (q_i,p_i)
\end{equation}
\begin{equation} \label{eq:79}
dW'_1 (q_i,p_i) = {\frac {\alpha} {\omega' |{\bf p}_i|}} \eta^2 [{\frac {1} {2}} (1 + \delta^2)(2 + {\frac {u_{a}^2} {1 + u_{a}}}) - {\frac {\gamma_i^2} {\eta^2}}] \cdot d^3 k'
\end{equation}
\begin{equation} \label{eq:80}
\begin{aligned}
d\sigma (p_f,q_i) = 2 Z^2 r_e^2 {\frac {|{\bf p}_f|} {|{\bf q}_i|}} {\frac {m^2 (m^2 + E_f q_{i0} + {\bf p}_f {\bf q}_i)} {[{\bf p}_f - {\bf q}_i + (l+1){\bf k}]^4}} \cdot \\ \cdot |L_{1+l}(p_f,q_i)|^2 d\Omega_f
\end{aligned}
\end{equation}
and
\begin{equation} \label{eq:81}
d\sigma_{(a)res} = dW'_1 (p_f,q_f) \cdot {\frac {2 \pi m |{\bf p}_f|} {|q_f^2 - m^2|^2}} \cdot d\sigma (q_f,p_i)
\end{equation}
\begin{equation} \label{eq:82}
\begin{aligned}
d\sigma (q_f,p_i) = 2 Z^2 r_e^2 {\frac {|{\bf q}_f|} {|{\bf p}_i|}} {\frac {m^2 (m^2 + q_{f0} E_f + {\bf q}_f {\bf p}_i)} {[{\bf q}_f - {\bf p}_i + (l+1){\bf k}]^4}} \cdot \\ \cdot |L_{1+l}(q_f,p_i)|^2 d\Omega_f
\end{aligned}
\end{equation}
\begin{equation} \label{eq:83}
dW'_1 (p_f,q_f) = {\frac {e^2 m^2} {\omega' |{\bf q}_f|}} \eta^2 [{\frac {1} {2}} (1 + \delta^2)(2 + {\frac {u_{b}^2} {1 + u_{b}}}) - {\frac {\gamma_f^2} {\eta^2}}] \cdot d^3 k'
\end{equation}
The ratios (18) and (19) establish the parameters $\gamma_i$ and $\gamma_f$ from the equations (79), (83) and propose the substitutions: $p_1 \rightarrow p_i$, $p_2 \rightarrow p_f$ to retrieve $\gamma_i$ and $p_1 \rightarrow q_f$, $p_2 \rightarrow p_f$ in extent to generate $\gamma_f$. Nevertheless, the relativistically invariant factors $u_{(a)}$ and $u_{(b)}$ are equal to:
\begin{equation} \label{eq:84}
u_{(a)} = {\frac {(kk')} {(kq_i)}} \approx {\frac {x'_{(a)}} {1 - x'_{(a)}}}, \: u_{(b)} = {\frac {(kk')} {(kp_f)}} \approx {\frac {x'_{(b)}} {1 - x'_{(b)}}}
\end{equation}
Proportions (78)-(80) and (81)-(83) indicate that for the channels A and B the RSB differential cross-sections effectively split into two first order processes with accordance to the fine structure constant. Therefore, for the channel A the foremost reaction is the laser-stimulated Compton effect on the initial electron ($dW'_1 (q_i,p_i)$ - is the probability of the phenomenon in a single unit of time [15]) and the sequential is the laser-modified Mott process of scattering of an intermediate electron on a nucleus ($d\sigma (p_f,q_i)$ - is the complementary differential cross-section [8, 11]). Furthermore, the channel B illustrates the analogous development of the resonant interaction. However, for the second alternative the laser-modified scattering of the initial electron ($d\sigma (q_f,p_i)$ - is the consonant differential scattering cross-section) initiates the system with consecutive laser-stimulated Compton emission of a spontaneous photon by the intermediate electron ($dW'_1 (p_f,q_f)$ - is the probability of the occurrence in a single unit of time). \\
The progression of the research realization postulates the reorganization of the relativistic resonant cross-sections (78) and (81) to harmonize the theoretical conception with the ambience of the resonant kinematics (35)-(39).
\begin{equation} \label{eq:85}
|L_{1+l}(p_f,q_i)|^2 = J_{1+l}^2 (\gamma_{fi}) \approx 1 \: \: (\gamma_{fi} \sim \eta \ll 1, l = -1)
\end{equation}
An equivalent expression coordinates for $|L_{1+l}(q_f,p_i)|^2 \approx 1$. To summarize, within the process of the (initial or intermediate) electron scattering on a nucleus the practically eminent effect is the mechanism with absorption of a single photon from the electromagnetic wave: $l = -1$.\\
The construct modeling of the phenomenon selects a circular polarization option $(\delta^2 = 1)$. For the resection of the resonant infinity in the channels A and B the investigation consolidates the method of the imaginary addition to the mass of the intermediate electron. Thus, for the channel A:
\begin{equation} \label{eq:86}
m \rightarrow \mu = m + i\Gamma_{(a)}, \: \Gamma_{(a)} = {\frac {q_{i0}} {2m}} W_{(a)}
\end{equation}
Where $W_{(a)}$ - is the total probability (per unit of time) of the laser-stimulated Compton effect on the intermediate electron with 4-momentum $q_i$ [15, 46].
\begin{equation} \label{eq:87}
W_{(a)} = {\frac {\alpha m^2} {4 E_i}} \eta \cdot K_i
\end{equation}
\begin{equation} \label{eq:88}
K_i = (1 - {\frac {4} {\varepsilon_i}} - {\frac {8} {\varepsilon_i^2}}) \ln (1+\varepsilon_i) + {\frac {1} {2}} + {\frac {8} {\varepsilon_i}} - {\frac {1} {2(1+\varepsilon_i)^2}}
\end{equation}
With application of the (87) expression the radiation width (86) acquires form:
\begin{equation} \label{eq:89}
\Gamma_{(a)} = {\frac {1} {8}} \alpha \eta^2 m K_i (1 - x'_{(a)})
\end{equation}
From the ratios (86)-(89) reexamination the resonant denominator realizes the following pattern:
\begin{equation} \label{eq:90}
|q_i^2 - \mu^2|^2 = m^4 [x'^2_{(a)} (\delta'^2_i - {\delta'^2_{(a)i}})^2 + {\frac {4 \Gamma^2_{(a)}} {m^2}}]
\end{equation}
where parameter $\delta'^2_{(a)i}$ is proportional to the resonant frequency of the spontaneous photon for the channel A from the statement (40).
\begin{equation} \label{eq:91}
\delta'^2_{(a)i} = {\frac {\varepsilon_i - (1 + \varepsilon_i) x'_{(a)}} {x'_{(a)}}}
\end{equation}
The resonant denominator for the channel B ascertains a similar dependency:
\begin{equation} \label{eq:92}
|q_f^2 - \mu^2|^2 = m^4 [x'^2_{(b)} (1 - x'_{(b)})^2 (\delta'^2_f - \delta'^2_{(b)f})^2 + {\frac {4 \Gamma^2_{(b)}} {m^2}}]
\end{equation}
Where
\begin{equation} \label{eq:93}
\Gamma_{(b)} = {\frac {1} {8}} \alpha \eta^2 m K_i
\end{equation}
and the parameter $\delta'^2_{(b)f}$ connects to the resonant frequency of the spontaneous photon of the channel B with (43) equation.
\begin{equation} \label{eq:94}
\delta'^2_{(b)f} = {\frac {\varepsilon_i - (1 + \varepsilon_i) x'_{(b)}} {x'_{(b)} (1 - x'_{(b)})^2}}
\end{equation}
As a result, the RSB differential cross-sections of the scattering of ultrarelativistic electrons for the pair of the channels of interaction (78) and (81) attain the configuration:
\begin{equation} \label{eq:95}
\begin{aligned}
d\sigma_{(a)res} =  {\frac {4 \pi^2 Z^2 \eta^2 \alpha r_e^2} {d^2(x'_{(a)})}} {\frac {(1-x'_{(a)})^2 \cdot D(x'_{(a)})} {[(\delta{'}^2_i - \delta{'}^2_{(a)i})^2 + \Gamma_{\delta_i}^2]}} \cdot \\ \cdot {\frac {dx{'}_{(a)}} {x{'}_{(a)}}} d{\delta'_i}^2 \cdot d{\delta'_f}^2 \cdot d\varphi_{-}
\end{aligned}
\end{equation}
\begin{equation} \label{eq:96}
\begin{aligned}
d\sigma_{(b)res} =  {\frac {4 \pi^2 Z^2 \eta^2 \alpha r_e^2} {d^2(x'_{(b)})}} {\frac {(1-x'_{(b)})^{-2} \cdot D(x'_{(b)})} {[(\delta{'}^2_f - \delta{'}^2_{(b)f})^2 + \Gamma_{\delta_f}^2]}} \cdot \\ \cdot {\frac {dx{'}_{(b)}} {x{'}_{(b)}}} d\delta{'}_i \cdot d\delta{'}_f \cdot d\varphi_{-}
\end{aligned}
\end{equation}
Where $\varphi_{-}$ - is the angle between the plane areas $({\bf k'},{\bf p}_i)$ and $({\bf k'},{\bf p}_f)$; $\Gamma_{\delta_i}$ and $\Gamma_{\delta_f}$ - are the angular radiation widths of the resonances for the A and B reaction schemes.
\begin{equation} \label{eq:97}
\Gamma_{\delta_i} =  {\frac {1} {4}} \alpha \eta^2 K_i ({\frac {1 - x'_{(a)}} {x'_{(a)}}}), \: \Gamma_{\delta_f} =  {\frac {1} {4}} \alpha \eta^2 {\frac {K_i} {x'_{(b)}(1-x'_{(b)})}}
\end{equation}
\begin{equation} \label{eq:98}
D(x')= 1 + (1-x')^2 - {\frac {4x'} {\varepsilon_i}} (1-{\frac {x'} {\varepsilon_i(1-x')}})
\end{equation}
\begin{equation} \label{eq:99}
d(x')= d_0 + ({\frac {m} {2 E_i}})^2 [d_1^2(x') + {\frac {\varepsilon_i} {\sin (\theta'/2)}} (\varepsilon_i +d_1(x'))]
\end{equation}
\begin{equation} \label{eq:100}
d_0 = \tilde{\delta'}^2_f + \delta_i^2 - 2 \delta_i \tilde{\delta'}_f \cos (\varphi_{-}), \: d_1(x') = (1 + \delta_i^2) - {\frac {(1 + \tilde{\delta'}^2_f)} {(1-x')}}
\end{equation}
\begin{equation} \label{eq:101}
\tilde{\delta'}_f = (1 - x') \delta'_f
\end{equation}
For the equal frequency range, however, in the absence of the external laser field the differential cross-section of the spontaneous bremsstrahlung process arranges into formation [46]:
\begin{equation} \label{eq:102}
\begin{aligned}
d\sigma_{0} = {\frac {1} {\pi}} Z^2 \alpha r_e^2 (1-x')^3 {\frac {[D_0(x') + (m/E_i)^2 D_1(x')]} {[d_0 + (m/2E_i)^2 d_1^2(x')]^2}} \cdot \\ \cdot {\frac {dx'} {x'}} d{\delta'_i}^2 d{\delta'_f}^2 \cdot d\varphi_{-}
\end{aligned}
\end{equation}
\begin{equation} \label{eq:103}
\begin{aligned}
D_0(x') = {\frac {\delta'^2_i} {(1 + \delta'^2_i)^2}} + {\frac {\tilde{\delta}'^2_f} {(1 + \tilde{\delta}'^2_f)^2}} + {\frac {x{'}^2} {2 (1 - x')}} \cdot \\ \cdot {\frac {(\delta'^2_i + \tilde{\delta}'^2_f)} {(1 + \delta'^2_i)(1 + \tilde{\delta}'^2_f)}} - [(1 - x') + {\frac {1} {(1 - x')}}] \cdot \\ \cdot {\frac {\delta'_i \tilde{\delta}'_f} {(1 + \delta'^2_i)(1 + \tilde{\delta}'^2_f)}} \cos \varphi_{-}
\end{aligned}
\end{equation}
\begin{equation} \label{eq:104}
D_1(x') = b_i(x') + {\frac {b_f(x')} {(1-x')^2}}
\end{equation}
\begin{equation} \label{eq:105}
b_i(x') =  {\frac {\delta'^2_i} {12(1+\delta'^2_i)^3}} \xi_i
\end{equation}
\begin{equation} \label{eq:106}
\begin{aligned}
\xi_i = [(1 - x') + {\frac {1} {(1 - x')}}] (9 + 4\delta'^2_i +3\delta'^4_i) - \\ - 2(1-\delta'^2_i)(3-\delta'^2_i) - {\frac {x'^2} {(1 - x')}} (9 + 2 \delta'^2_i + \delta'^4_i)
\end{aligned}
\end{equation}
The expression for $b_f(x')$ develops as a deriver from the (105), (106) proportions with substitution of ${\delta'_i}^2 \rightarrow {\tilde{\delta'_f}}^2$. It is important to emphasize that the differential cross-sections (95), (96) and (102) compile the small-scaled approximations that coordinate to the level of $\sim (m/E_i)^2 \ll 1$ and generate the dominant contribution to the differential cross-section degree under the limitations of $\delta'_i \sim \tilde{\delta'}_f \gtrsim 1$ and
\begin{equation} \label{eq:107}
\varphi_{-} \lesssim {\frac {m} {E_i}}, \: |\delta'_i - \tilde{\delta'}_f| \lesssim {\frac {m} {E_i}}
\end{equation}
With specified prerequisites the magnitudes $D_0(x') \rightarrow 0$, $d_0 \rightarrow 0$ and, additionally, the according differential cross-sections achieve the sharp maximums. Therefore, the differential cross-sections without the field (102) and in the field (95), (96) in the kinematical region (107) obtain the following order:
\begin{equation} \label{eq:108}
d\sigma_{0} \sim Z^2 \alpha r_e^2 ({\frac {E_i} {m}})^2, \: d\sigma_{(a)res} \lesssim d\sigma_{0} ({\frac {E_i} {m}})^2 (\alpha \eta)^{-2}
\end{equation}
After integration of the resonant cross-sections (95), (96), and the cross-section in the absence of the field (102) on the azimuthal angle $\varphi_{-}$ with supplementary calculations the research establishes:
\begin{equation} \label{eq:109}
\begin{aligned}
d\sigma_{(a)res} = 8 \pi^3 Z^2 \alpha r_e^2 \eta^2 {\frac {\delta'^2_i + \tilde{\delta}'^2_f} {[b(x{'}_{(a)}) \cdot G(x{'}_{(a)})]^{3/2}}} \cdot \\ \cdot {\frac {(1-x'_{(a)})^2 \cdot D(x'_{(a)})} {[(\delta'^2_i - \delta'^2_{(a)i})^2 + \Gamma_{\delta_i}^2]}} {\frac {dx{'}_{(a)}} {x{'}_{(a)}}} d\delta'^2_i \cdot d\delta'^2_f
\end{aligned}
\end{equation}
\begin{equation} \label{eq:110}
\begin{aligned}
d\sigma_{(b)res} = 8 \pi^3 Z^2 \alpha r_e^2 \eta^2 {\frac {\delta'^2_i + \tilde{\delta}'^2_f} {[b(x{'}_{(b)}) \cdot G(x{'}_{(b)})]^{3/2}}} \cdot \\ \cdot {\frac {(1-x'_{(b)})^{-2} \cdot D(x'_{(b)})} {[(\delta'^2_f - \delta'^2_{(b)f})^2 + \Gamma_{\delta_f}^2]}} {\frac {dx{'}_{(b)}} {x{'}_{(b)}}} d\delta'^2_i \cdot d\delta'^2_f
\end{aligned}
\end{equation}
\begin{equation} \label{eq:111}
d\sigma_{0} = 4 Z^2 \alpha r_e^2 {\frac {(1-x')^3} {[b(x')]^{3/2}}} (\delta'^2_i + \tilde{\delta}'^2_f) \cdot D_2(x') \cdot {\frac {dx'} {x'}} d\delta'^2_i d\delta'^2_f
\end{equation}
Where:
\begin{equation} \label{eq:112}
G(x') = 1 +  ({\frac {m} {E_i}})^2 {\frac {\varepsilon_i \cdot (\delta'^2_i + \tilde{\delta}'^2_f)} {2 \sin^2(\theta'/2)}} {\frac {[\varepsilon_i + d_1 (x')]} {b(x')}}
\end{equation}
\begin{equation} \label{eq:113}
b(x') = (\delta'^2_i - \tilde{\delta}'^2_f)^2 + {\frac {1} {2}} ({\frac {m} {E_i}})^2 (\delta'^2_i + \tilde{\delta}'^2_f) \cdot d_1^2(x')
\end{equation}
\begin{equation} \label{eq:114}
D_2(x') = D'_0(x') + (m/E_i)^2 D'_1(x')
\end{equation}
\begin{equation} \label{eq:115}
\begin{aligned}
D'_0(x') = {\frac {\delta'^2_i} {(1 + \delta'^2_i)^2}} + {\frac {\tilde{\delta}'^2_f} {(1 + \tilde{\delta}'^2_f)^2}} + {\frac {x{'}^2} {2 (1 - x')}} \cdot \\ \cdot {\frac {(\delta'^2_i + \tilde{\delta}'^2_f)} {(1 + \delta'^2_i)(1 + \tilde{\delta}'^2_f)}} - [(1 - x') + {\frac {1} {(1 - x')}}] \cdot \\ \cdot {\frac {2 \delta'_i \tilde{\delta}'_f} {(1 + \delta'^2_i)(1 + \tilde{\delta}'^2_f)(\delta'^2_i + \tilde{\delta}'^2_f)}}
\end{aligned}
\end{equation}
\begin{equation} \label{eq:116}
\begin{aligned}
D_1' (x') = D_1(x')+ [(1 - x') + {\frac {1} {(1 - x')}}] \cdot \\ \cdot {\frac {d_1^2 \delta'_i \tilde{\delta}'_f} {2 (1 + \delta'^2_i)(1 + \tilde{\delta}'^2_f)(\delta'^2_i + \tilde{\delta}'^2_f)}}
\end{aligned}
\end{equation}
The resonant denominators of the expressions (109), (110) represent a characteristic Breit-Wigner form. Within the conditions $\delta'^2_i \rightarrow \delta'^2_{(a)i}$ (for channel A) and $\delta'^2_f \rightarrow \delta'^2_{(b)f}$ (for channel B) the resonant aspects of the phenomenon reach substantial actualization. Consequently, statements (109) and (110) promote the simulation of the coordinate evaluations and the investigation estimates the proportions for the maximal differential cross-sections in the units of the cross-section without the external field (111):
\begin{equation} \label{eq:117}
R_{(j)res}^{max} = {\frac {d\sigma_{(j)res}^{max}} {d\sigma_0}} = f_0 \cdot R_{(j)}, \: f_0 = {\frac {32 \pi^3} {\eta^2 \alpha^2}},
\end{equation}
\begin{equation} \label{eq:118}
R_{(j)} = {\frac {x'^2_{(j)}} {(1-x'_{(j)})^3}} \cdot {\frac {D(x'_{(j)})} {D_2(x'_{(j)})}} \cdot {\frac {[G(x'_{(j)})]^{-3/2}} {K_i^2}}, \: j = a,b
\end{equation}
Equations (117) and (118) define the RSB differential cross-section (in the units of the differential cross-section in the absence of the laser field) for the channels A and B with simultaneous registration of the emission angles of the final electron and spontaneous photon (parameters $\delta'_i$ and $\delta'_f$) and the spontaneous photon frequencies within the diapason from $\omega'_{(a)}$ to $[\omega{'}_{(a)} + d\omega{'}_{(a)}]$ (for the channel A) and from $\omega{'}_{(b)}$ to $[\omega{'}_{(b)} + d\omega{'}_{(b)}]$ (for the channel B). In addition, the spontaneous photon radiation angle with interconnection to the momentum of the initial electron (parameter $\delta'^2_i$) systematizes the resonance frequency $\omega'_{(a)}$ (40) and the energy of the final electron $E_f \approx E_i - \omega{'}_{(a)}$. Moreover, the indicated magnitudes do not depend on the angle of the final electron emission (parameter $\delta{'}^2_f$). Contrastingly, interactions in the channel B delineate the alternative framework. Thus, the radiation angle of the spontaneous photon in compound to the final electron momentum (parameter $\delta{'}^2_f$) conducts the resonance frequency $\omega{'}_{(b)}$ and the energy of the final electron $E_f \approx E_i - \omega{'}_{(b)}$ (see (54)-(61)). Accordingly, the parameter $\delta'^2_i$ does not specify the designated physical values. \\
The resonant frequencies from the $R_{(a)}$ and $R_{(b)}$ for the pair of the reaction channels A and B represent an essentially segregated attitude. The functions $R_{(a)}$, $R_{(b)}$ and $f_0$ allocate the magnitude of the maximal resonant cross-section (see (117), (118)). The resonance radiation width generally designates the $f_0$ and the value of it particularizes a considerable degree: for the laser wave intensity of  $\eta = 0.1$ $(I \sim (10^{16} \div 10^{17} \: W/cm^2)$ the function ${f_0} \approx 1.86 \cdot 10^9$. Within the experiment realization the resonance width is substantially momentous in comparison to the radiational width and the function $f_0$ is moderate.
Alternatively, the transmitted momentum within the limitation of (107) fulfillment structures the magnitudes of functions $R_{(a)}$ and $R_{(b)}$. The Fig. 5 illustrates the dependency for the channel A of the function $R_{(a)}$ from the parameter $\delta'^2_i$ for the electron energies $E_i \approx 167 GeV$ (Fig. 5a) and $E_i \approx 250 GeV$ (Fig. 5b) for the fixed parameter $\delta{'}^2_f$ values. The plot demonstrates that the function $R_{(a)}$ obtains 6-8 orders exceeding surplus within the resonance when the parameters $\delta{'}^2_i = \tilde{\delta}'^2_f$ coincide. The Fig. 6 for the channel B highlights the function $R_{(b)}$ relation on the parameter $\tilde{\delta}'^2_f$ for the electron energies $E_i \approx 167 GeV$ (Fig. 6a) and $E_i \approx 417 GeV$ (Fig. 6b) for the definitive parameter $\delta'^2_i$ rate. The graph contour clarifies that with contemporaneous parameters $\tilde{\delta}'^2_f = \delta{'}^2_i$ degree the function $R_{(b)}$ surmounts by 8-11 orders the standard ratio within the resonance environment.

\begin{figure}[h!]
   \begin{minipage}{.5\textwidth}
       \includegraphics[width=7cm]{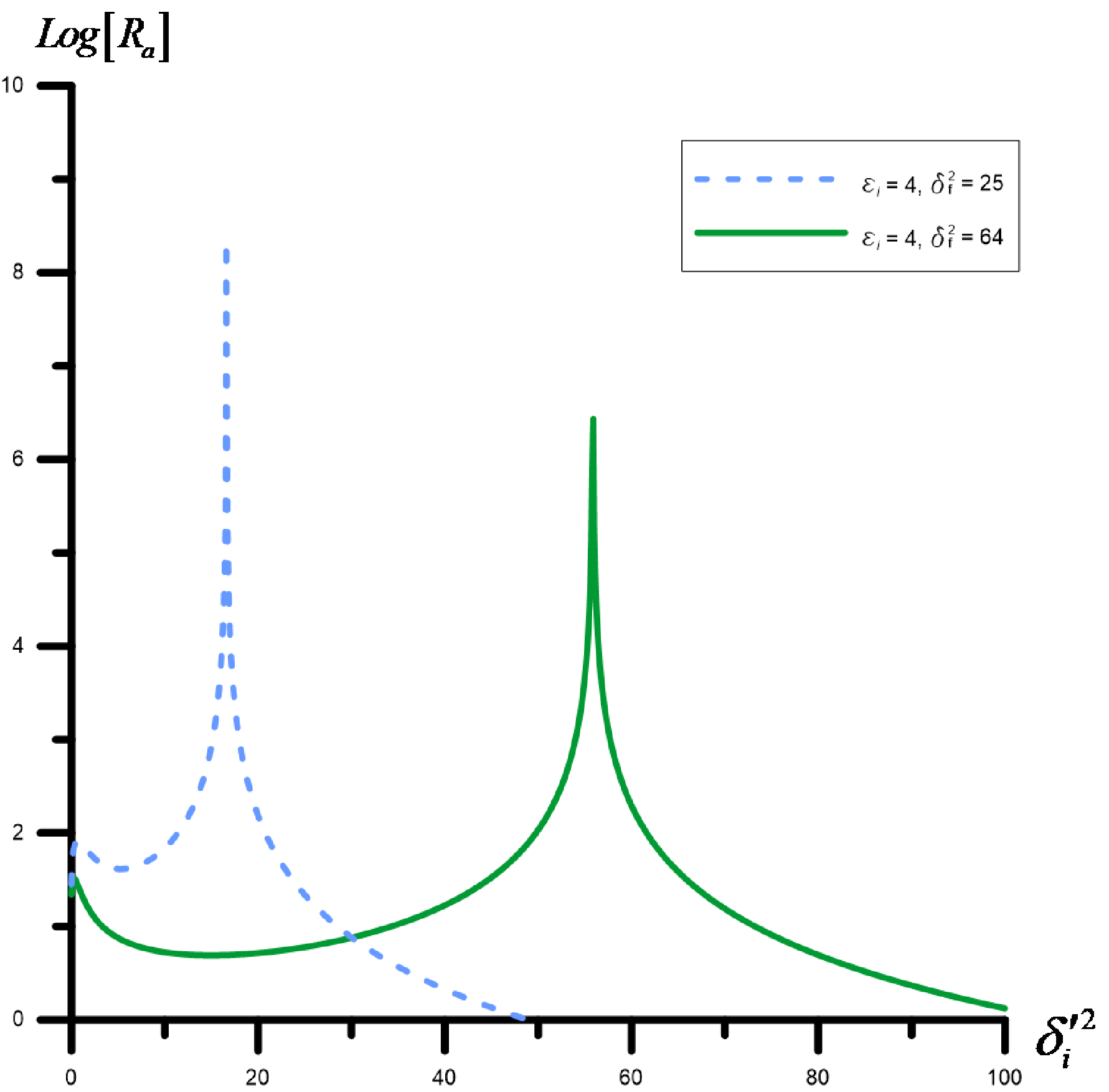}
  \end{minipage}
  \begin{minipage}{.5\textwidth}
      \includegraphics[width=7cm]{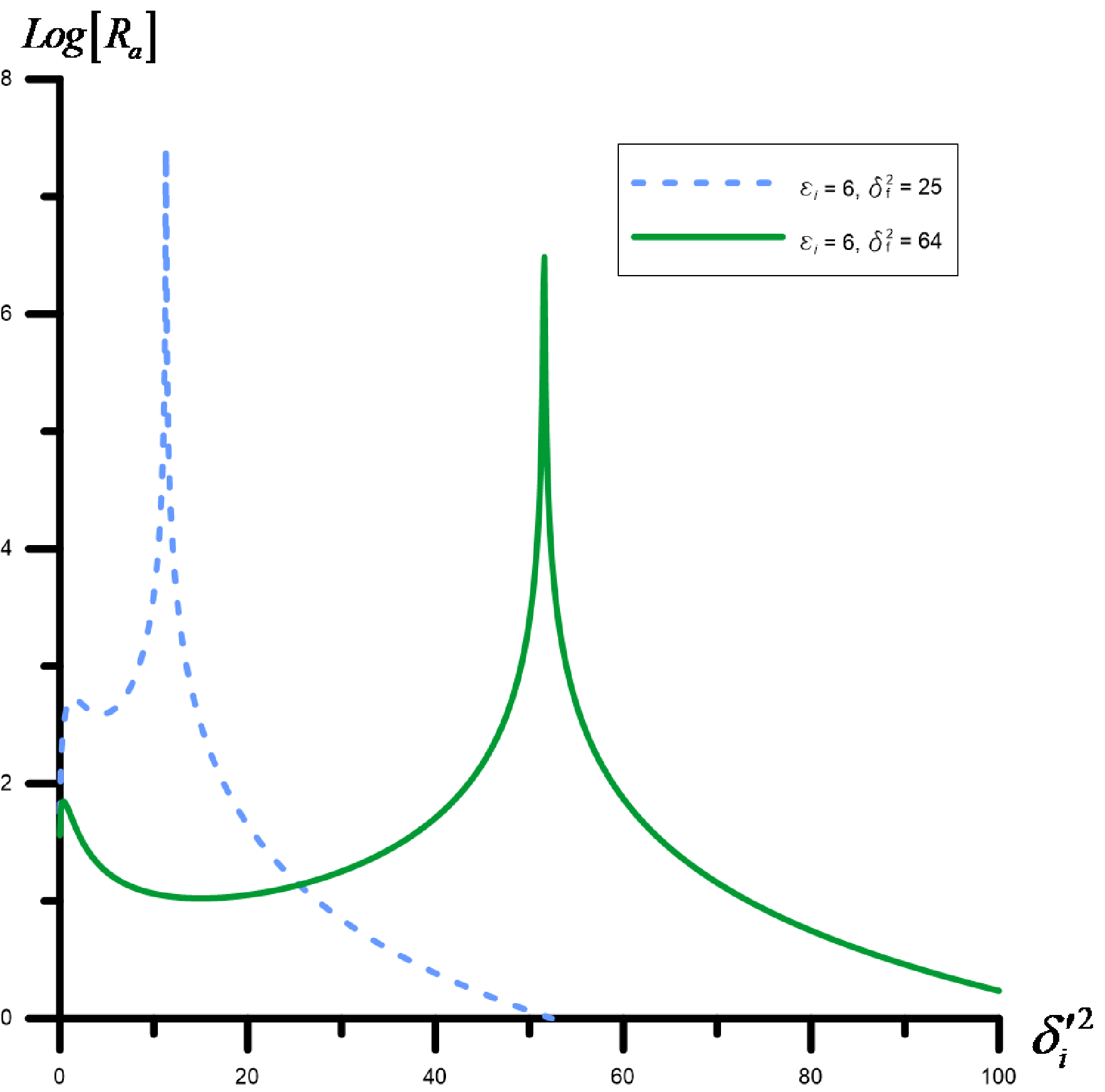}
  \end{minipage}
  \caption{Dependency of the RSB differential cross-section for the ultrarelativistic electrons on the parameter $\delta{'}^2_i$ for the channel A (117), (118), (40) for the fixed radiation angles of the final electron (parameter $\delta{'}^2_f$). Fig. 5a - delineates the electron energies $E_i \approx 167 GeV$, Fig. 5b - designates the electron energies $E_i \approx 250 GeV$.}
  \label{Fig:Figure 5}
\end{figure}

\begin{figure}[h!]
   \begin{minipage}{.5\textwidth}
       \includegraphics[width=7cm]{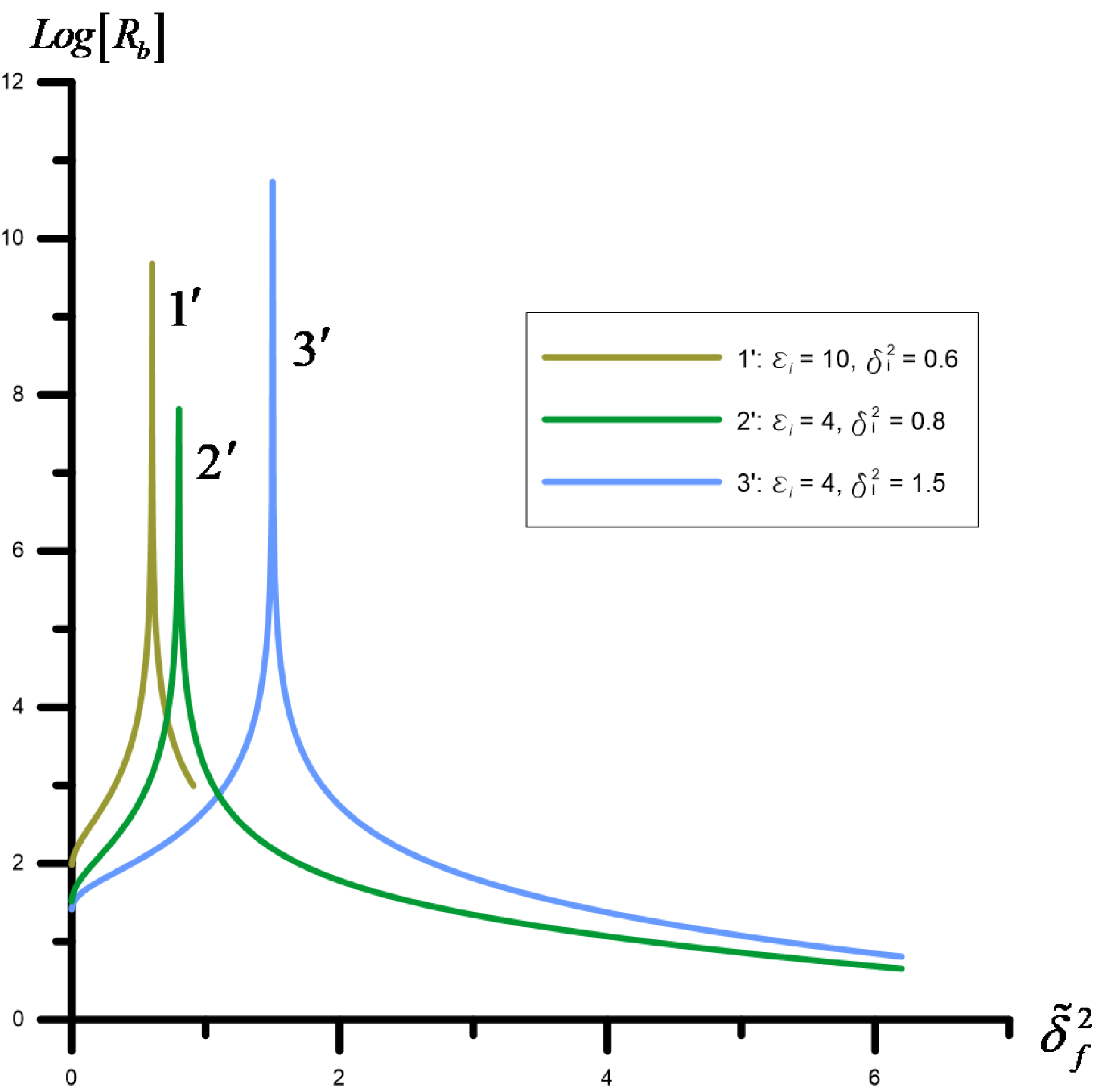}
  \end{minipage}
  \begin{minipage}{.5\textwidth}
      \includegraphics[width=7cm]{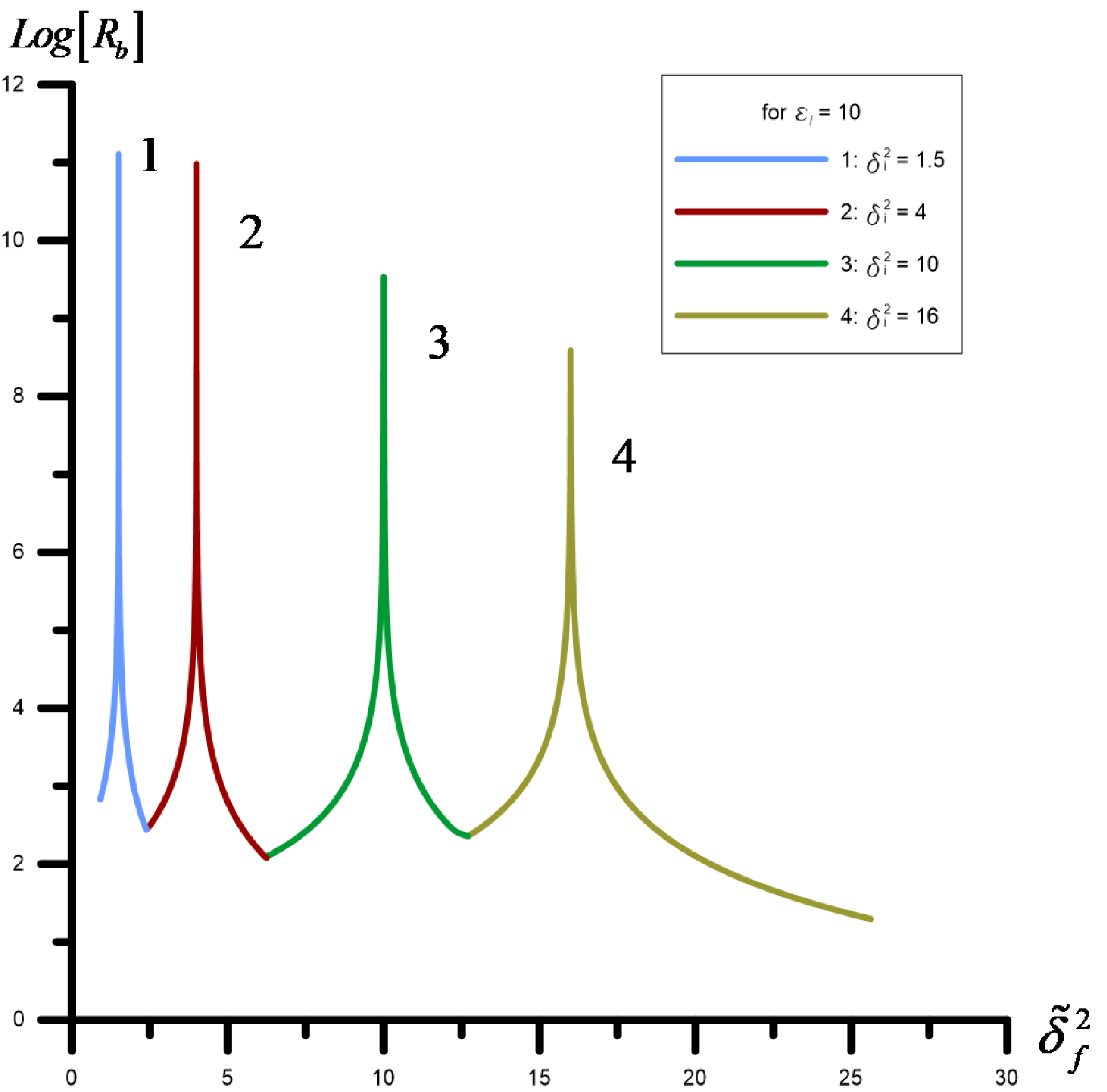}
  \end{minipage}
  \caption{Dependency of the RSB differential cross-section for the ultrarelativistic electrons on the parameter $\tilde{\delta}'^2_f$ for the channel B (117), (118), (54), (59) for the definitive parameter $\delta'^2_i$ magnitudes. Sharp peaks $2',3'$ characterize the electron energies $E_i \approx 167 GeV$ and the parameter $\delta'^2_f$ fluctuation within the interval (60). Peak $1'$ describes the electron energy $E_i \approx 417 GeV$ and the parameter $\delta'^2_f$ alteration in the $0 < \delta'^2_f \leqslant \delta'^2_{-}$ spectrum. Peaks $1,2,3$ , delineate the parameter $\delta'^2_f$ dispersion within the range of $\delta'^2_{-} < \delta'^2_f \leqslant \delta'^2_{+}$ and specify three various frequency possibilities (54), peak $4$ - designates the parameter $\delta'^2_f$ modification in the diapason $\delta'^2_{+} < \delta'^2_f$ for the electron energy $E_i \approx 417 GeV$.}
  \label{Fig:Figure 6}
\end{figure}

\noindent Subsequently, the investigation development proposes integration of the resonant differential cross-section for the channel A (109) on the parameter $\delta'^2_f$, and for the channel B on the parameter $\delta'^2_i$:
\begin{equation} \label{eq:119}
\begin{aligned}
d\sigma_{(a)res} = 32 \pi^3 Z^2 \alpha r_e^2 ({\frac {E_{*}} {m}})^2 \cdot {\frac {\eta^2 D(x'_{(a)})} {[(\delta'^2_i - \delta'^2_{(a)i})^2 + \Gamma_{\delta_i}^2]}} \cdot \\ \cdot {\frac {dx{'}_{(a)}} {x{'}_{(a)}}} d\delta'^2_i
\end{aligned}
\end{equation}
\begin{equation} \label{eq:120}
\begin{aligned}
d\sigma_{(b)res} = 32 \pi^3 Z^2 \alpha r_e^2 ({\frac {E_{*}} {m}})^2 \cdot {\frac {\eta^2 (1 - x'_{(b)})^{-2} D(x'_{(b)})} {[(\delta'^2_f - \delta'^2_{(b)f})^2 + \Gamma_{\delta_f}^2]}} \cdot \\ \cdot {\frac {dx{'}_{(b)}} {x{'}_{(b)}}} d\delta'^2_f
\end{aligned}
\end{equation}
\\
When $\delta'^2_i \rightarrow \delta'^2_{(a)i}$ (for the channel A) and $\delta'^2_f \rightarrow \delta'^2_{(b)f}$ (for the channel B) the resonant differential cross-sections (119) and (120) attain maximums:
\begin{equation} \label{eq:121}
{\frac {d{\sigma'^{max}_{(a)res}}} {d{\delta'}^2_i dx'_{a}}} = (Z^2 \alpha r_e^2) g_0 \cdot F_{(a)}
\end{equation}
\begin{equation} \label{eq:122}
{\frac {d{\sigma'^{max}_{(b)res}}} {d{\delta'}^2_f dx'_{b}}} = (Z^2 \alpha r_e^2) g_0 \cdot F_{(b)}
\end{equation}
\begin{equation} \label{eq:123}
g_0 = {\frac {512 \pi^3} {(\alpha \eta)^2}} ({\frac {E_{*}} {m}})^2
\end{equation}
Equations (121), (122) indicate that the basis of the maximal resonant cross-section fundamentally depends on the function $g_0$ (123) that in the range of the optical frequencies for the electromagnetic wave intensity $\eta = 0.1$ $(I \sim (10^{16} \div 10^{17} \: W/cm^2)$ and characteristic energy $E_{*} = 41.7 GeV$ of the process is equal to:
\begin{equation} \label{eq:124}
g_0 \approx 2 \cdot 10^{20}
\end{equation}
Nevertheless, the approximately considerable magnitude of the function $g_0$ coordinates with the small-scaled radiation width of the resonance that apportions the augmentation of an order $\sim 10^6 \div 10^8$ and with minimal transferred momenta of the resonant phenomenon which primary affect the $g_0$ physical quantity.
\\
\\
\\
\\
\\
\\
\\
\\
\\
\\
\\
\\
\\
\\
\\
\\
\\

\end{document}